\documentclass[11pt,a4paper]{article}
\pdfoutput=1

\usepackage[colorlinks=true, linkcolor=black!50!blue, urlcolor=blue, citecolor=blue, anchorcolor=blue]{hyperref}
\usepackage[font=small,labelfont=bf,labelsep=period]{caption}
\usepackage[a4paper,top=2.5cm,bottom=2.5cm,left=2.5cm,right=2.5cm,bindingoffset=0mm]{geometry}

\usepackage{graphicx}
\usepackage{float}
\usepackage{afterpage}
\usepackage{epsfig}
\usepackage{cite}
\usepackage{amssymb}
\usepackage{amsmath}
\usepackage{dsfont}
\usepackage{multirow}
\usepackage{url}
\usepackage[table]{xcolor}
\usepackage{afterpage}
\usepackage{hyperref}
\usepackage{booktabs}
\usepackage{url}
\usepackage{hyperref}
\usepackage{tabularx}
\usepackage{breqn}
\usepackage{tikz}

\usetikzlibrary{shapes,arrows}

\tikzstyle{block}=[rectangle, draw, fill=blue!20, text width=9.5em, 
                   text centered, rounded corners, minimum height=2em, 
                   minimum width=10em]
\tikzstyle{line} =[draw, -latex']
\tikzstyle{decision} = [diamond, draw, fill=red!20, text width=7.5em, text centered,  inner sep=0pt, minimum height=2em, aspect=4]
\tikzstyle{cloud} = [draw, ellipse,fill=green!20, minimum height=2em]
\tikzstyle{inout} = [rectangle, draw, fill=green!20, text width=9.5em, text centered, rounded corners, minimum height=2em, minimum width=10em]

\newcommand{\smallfrac}[2]{{\mathchoice{\textstyle{\frac{#1}{#2}}}%
  {\scriptstyle{\frac{#1}{#2}}}{\scriptscriptstyle{\frac{#1}{#2}}}{#1/#2}}}

\newcolumntype{C}{>{\centering\arraybackslash}X}

\begin{document}

\begin{flushright}
Edinburgh 2020/15\\
Nikhef/2020-033\\
\end{flushright}
\vspace{0.3cm}

\begin{center}
  {\Large \bf Deuteron Uncertainties in the Determination of Proton PDFs\\}
  \vspace{1.5cm}
  {\small
  Richard~D.~Ball$^1$, 
  Emanuele~R.~Nocera$^{1,2}$
  and Rosalyn~L.~Pearson$^1$}

\vspace{0.5cm}
{$^1${\it\small The Higgs Centre for Theoretical Physics,\\ 
 University of Edinburgh, JCMB, KB, Mayfield Rd, Edinburgh EH9 3FD, Scotland}\\
 $^2${\it\small Nikhef Theory Group, 
 Science Park 105, 1098 XG Amsterdam, The Netherlands}\\
}

\vspace{1cm}

{\bf \large Abstract}

\end{center}

We evaluate the uncertainties due to nuclear effects in global
fits of proton parton distribution functions (PDFs) that utilise
deep--inelastic scattering and Drell--Yan data on deuterium targets. To do 
this we use an iterative procedure to determine proton and deuteron 
PDFs simultaneously, each including the uncertainties in the other.
We apply this procedure to determine the nuclear uncertainties in the
SLAC, BCDMS, NMC and DYE866/NuSea fixed target deuteron data included
in the NNPDF3.1 global fit. We show that the effect of the nuclear uncertainty
on the proton PDFs is small, and that the increase in overall uncertainties is
insignificant once we correct for nuclear effects.
\newline

Parton distribution functions (PDFs) are an essential ingredient in the 
theoretical predictions of hadronic observables at the LHC~\cite{Gao_2018,
  Ethier:2020way,Forte:2020yip}. PDFs for the proton are determined via global
QCD fits to a range of experimental data, including those where the proton is
not in a free state. In particular, these include deep--inelastic scattering
(DIS) and Drell--Yan (DY) fixed target collisions involving deuterium and heavy
nuclear  targets. In these processes the interaction of the proton is altered
due to nuclear effects, and this difference propagates through to the fitted
PDFs. Measurements involving deuterium targets still play a significant role
in the determination of proton PDFs, in particular to separate the up and down
flavours for large momentum fraction $x$,  a region which is especially
important for searches for physics beyond the Standard Model. Because of this,
deuteron corrections have been extensively studied and have been included in
PDF analyses via parametrizations of a nuclear smearing
function~\cite{Owens:2012bv,Ball:2013gsa,Harland-Lang:2014zoa,Accardi:2016qay,
  Alekhin:2017fpf}, inspired by various deuteron wavefunction
models~\cite{Wiringa:1994wb,Melnitchouk:1994rv,Melnitchouk:1996vp,
  Machleidt:2000ge,Gross:2014wqa}. This approach relies
on model assumptions, which can ultimately bias the determination of the
PDFs in a way which is difficult to quantify. Because the precision of the PDFs
is now constrained by the data to a few percent for most quark flavours in a
wide kinematic range~\cite{Ball:2017nwa}, a faithful
estimate~\cite{Ball:2018lag} of the theoretical uncertainty associated
with nuclear effects (potentially of comparable size) is becoming
necessary.

In a previous study~\cite{Ball:2018twp} we showed how theoretical uncertainties
due to heavy nuclear targets in DIS and DY measurements can be incorporated into
global fits of proton PDFs. Specifically, in the framework of the NNPDF
methodology (see~\cite{Ball:2014uwa} and references therein for a comprehensive
description), we added to the experimental covariance matrix a theoretical
covariance matrix, accounting for the additional uncertainties due to nuclear
effects. Two distinct procedures were adopted: in the first, the contribution
of the nuclear data to the PDF fit is deweighted by an uncertainty that
encompasses both the difference between proton and nuclear PDFs and the
uncertainty in the nuclear PDFs; in the second, the difference between proton
and nuclear PDFs is used to correct the theoretical predictions, while the
deweighting only takes the nuclear PDF uncertainty into account, and is
therefore correspondingly smaller. If the uncertainty in the nuclear PDFs is
correctly estimated, and smaller than the shift, the second procedure should
give more precise results. The nuclear PDFs were determined as an equally
weighted replica average of the DSSZ~\cite{deFlorian:2011fp},
nCTEQ15~\cite{Kovarik:2015cma}, and EPPS16~\cite{Eskola:2016oht} PDF sets
for the relevant heavy nuclei (Cu, Fe and Pb).
Despite the fact that these are obtained from a global analysis of experimental
data taken in a wide variety of processes, there are sizeable differences
between them. This suggests that these three sets might not be sufficiently
consistent to determine a precise nuclear correction, but can be used to
estimate the uncertainty due to nuclear effects, and indeed the second
procedure led to a worse global fit than the first.

Nuclear corrections to deuterium are rather smaller than those for heavy nuclei.
In the framework of the NNPDF methodology, these corrections have been studied
in a dedicated work~\cite{Ball:2013gsa}, based on the NNPDF2.3
release~\cite{Ball:2012cx}, and again in the context of the NNPDF3.0 and
NNPDF3.1 determinations (see, respectively, Sect.~5.1.4 in~\cite{Ball:2014uwa}
and Sect.~4.11 in~\cite{Ball:2017nwa}). Variants of the NNPDF2.3, NNPDF3.0 and
NNPDF3.1 fits were performed by correcting all deuterium data
according to Eq.~(8) of~\cite{Harland-Lang:2014zoa}, with parameter values
determined in~\cite{Martin:2012da}. In all cases results were consistent.
Specifically, for NNPDF3.1, it turned out that the central value of the up and
down quark PDFs were moderately affected at large $x$ (less than half
a sigma), and that the corresponding uncertainty was somewhat increased.
Other PDFs were hardly affected. A slight increase in the global $\chi^2$
was observed, a fact that suggested that the theoretical uncertainty associated
with the nuclear correction was not optimally quantified. For these reasons,
nuclear corrections were not included in the baseline NNPDF3.1 set.

In this paper, we revisit the impact of nuclear corrections in deuterium data
by extending the approach developed for heavy nuclei in~\cite{Ball:2018twp}.
We focus on the dataset included in the NNPDF3.1 PDF
determination~\cite{Ball:2018iqk}, which is made up of 3978 data points
(see~\cite{Ball:2017nwa} for details). Out of these, 418 data points (about
10\% of the whole dataset) come from experiments using deuterium targets,
specifically SLAC~\cite{Whitlow:1991uw}, BCDMS~\cite{Benvenuti:1989fm},
NMC~\cite{Arneodo:1996kd}, and DYE866/NuSea~\cite{Towell:2001nh}. The DIS data
are in the form of deuteron to proton structure function ratios, $F_2^d/F_2^p$,
for NMC, and of deuteron structure functions, $F_2^d$, for SLAC and BCDMS; the
DY data is in the form of ratios of cross sections for a proton beam on a
deuteron target to a proton beam on a proton target,
$\sigma^{\rm DY}_{pd}/\sigma^{\rm DY}_{pp}$, for DYE866/NuSea.

A significant weakness in our treatment of nuclear effects in heavy nuclei was
its dependence on externally determined nuclear PDFs. To avoid this when
treating deuterium, we fit our own deuterium PDFs directly from the deuteron
data by means of a procedure which is iterated to consistency with the global
proton fit. In this way we account simultaneously for the nuclear uncertainties
in the deuteron when determining global proton PDFs, and the uncertainties in
the proton PDF when determining the deuteron PDF (and thus the nuclear
correction). The advantage of the new approach is that the deuteron and proton
fits are all performed using a consistent theoretical and methodological
fitting framework, and the resulting nuclear corrections and their
uncertainties are thus equally reliable and unbiased. It is very similar to
the self consistent procedure set out in \cite{Ball:2018lag} for the
simultaneous determination of PDFs and fragmentation functions using
experimental data from semi--inclusive DIS. 

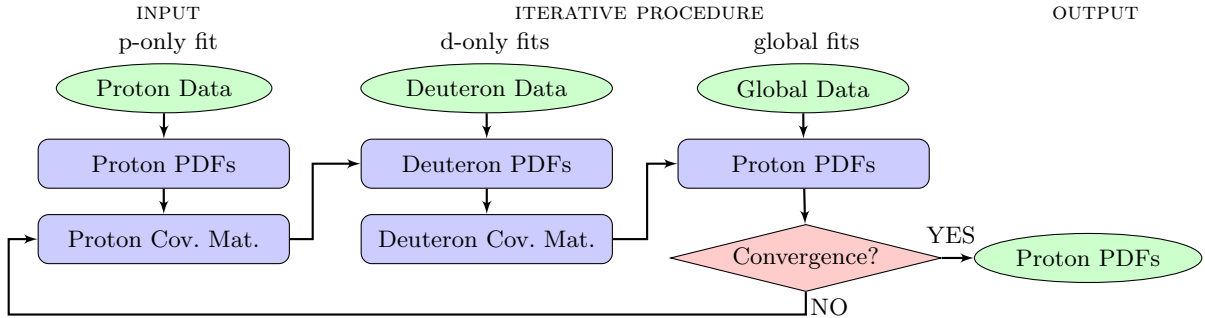
\begin{figure}[!t]
  \centering
  \footnotesize
  \begin{tikzpicture}[node distance = 1.4cm, auto]
    \node at(-6.6,2.5) {\sc input};
    \node at(+5.6,2.5) {\sc output};
    \node at(-0.4,2.5) {\sc iterative procedure};
    \node at(-6.6,2.1) {p-only fit};
    \node at(-2.3,2.1) {d-only fits};
    \node at(+1.8,2.1) {global fits};
    \node at(+2.1,-1.4) {NO};
    \node at(+3.7,-0.45) {YES};
    \node [cloud, left=5.23cm] at (0,1.5)  (pdata)   {Proton Data};
    \node [block, left=5cm] at (0,0.5)     (pPDFs0)  {Proton PDFs};
    \node [block, left=5cm] at (0,-0.5)    (pcovmat) {Proton Cov.~Mat.};
    \node [cloud, left=0.75cm] at (0,1.5)    (ddata)   {Deuteron Data};
    \node [block, left=0.75cm] at (0,0.5)  (dPDFs)   {Deuteron PDFs};
    \node [block, left=0.75cm] at (0,-0.5) (dcovmat) {Deuteron Cov.~Mat.};
    \node [cloud, left=-3.15cm] at (0,1.5)    (gdata)   {Global Data};
    \node [block, right=0.1cm] at (0,0.5)  (pPDFs)   {Proton PDFs};
    \node [decision, right=0cm] at (0,-0.75)  (conv)    {Convergence?};
    \node [cloud, right=4cm] at (0,-0.75)     (final)   {Proton PDFs};
    \path [line, thick] (pdata)  -- (pPDFs0);
    \path [line, thick] (pPDFs0) -- (pcovmat);
    \path [line, thick] (ddata)  -- (dPDFs);
    \path [line, thick] (pcovmat) -| (-4.7,-0.5) |- (dPDFs);
    \path [line, thick] (dPDFs) -- (dcovmat);
    \path [line, thick] (dcovmat) -- (-0.3,-0.5) |- (pPDFs);
    \path [line, thick] (pPDFs) -- (conv);
   \path [line, thick] (gdata)  -- (pPDFs);
   \path [line, thick] (conv) |- (0,-1.5) -- (-8.7,-1.5) |- (pcovmat);
    \path [line, thick] (conv) -- (final);
  \end{tikzpicture}
  \caption{A schematic representation of the iterative procedure adopted to
    determine the uncertainty due to deuteron corrections in proton PDF fits,
    see text for details. The global dataset is the union of the proton and deuteron datasets.}
  \label{fig:procedure}
\end{figure}

The logical structure of the procedure is depicted in Fig.~\ref{fig:procedure}.
The NNPDF3.1 global dataset is split into two disjoint 
subsets: `deuteron data', including the aforementioned datasets
(SLAC, BCDMS, NMC and DYE866/NuSea); and `proton
data' including all the other datasets. The global dataset is the union
of deuteron and proton datasets. Note that the proton data also includes
CHORUS~\cite{Onengut:2005kv}, NuTeV~\cite{Goncharov:2001qe},
and DYE605~\cite{Moreno:1990sf} data taken in experiments on heavy nuclei.
The effect of nuclear corrections on these measurements was studied
in~\cite{Ball:2018twp} and is not considered here as we want to focus
exclusively on the nuclear effects in deuterium.

The deuteron data themselves may be split into two sets: `pure' deuteron data,
from DIS on a deuteron target (the SLAC and BCDMS data for $F_2^d$), and
`mixed' deuteron data, which also involve protons (the NMC data for the ratio
$F_2^d/F_2^p$, and the DYE866/NuSea DY data for the ratio
$\sigma^{DY}_{pd}/\sigma^{DY}_{pp}$). We denote the theoretical predictions for the
pure deuteron data by $T_i^d[f_d]$, where $f_d$ is the deuteron PDF, and the
theoretical predictions for the mixed deuteron data by $T_i^d[f_d,f_p]$, where
$f_p$ is the proton PDF. In each case the index $i$ runs over the individual
data points. In a conventional global proton fit, without deuteron nuclear
corrections, the deuteron observables $T_i^d[f_d]$ and $T_i^d[f_d,f_p]$ are
included in the fit by replacing $f_d$ by the isoscalar PDF
\begin{equation}
  f_s \equiv \smallfrac{1}{2}(f_p+f_n).
  \label{eq:isoscalar}
\end{equation}
Here $f_n$ is the neutron PDF, determined from the proton PDF by assuming exact
isospin invariance (and thus in practice by swapping the up and down PDFs).

The proton data are used to determine a first set of pure proton PDFs
$\{f_p^{(k)}: k=1\cdots N_{\rm rep}\}$, using the usual NNPDF methodology.
The central prediction is $f_p^{(0)} = \langle f_p^{(k)}\rangle$, where the
angled brackets denote a simple average over the $N_{\rm rep}$ Monte Carlo
replicas. If the deuteron data were all pure deuteron data, in practice only
the SLAC and BCDMS datasets, we could simply produce a similar
set of pure deuteron PDFs $\{f_d^{(k)}: k=1\cdots N_{\rm rep}\}$.
These by construction will include the nuclear effects and the size of the
nuclear correction would be $T_i^d[f_d^{(0)}]-T_i^d[f_s^{(0)}]$,
with $f_s^{(0)}$ determined from the proton PDFs using
Eq.~\eqref{eq:isoscalar} averaged over proton PDF replicas.
To include the mixed deuteron data, in particular the data from NMC and
DYE866/NuSea, we have to be more careful since to evaluate the theoretical
predictions for a given deuteron PDF we also need  
a proton PDF. For this we can use the central value of the pure proton fit
$f_p^{(0)}$ (i.e. replica zero, the average of all the other replicas) and the
size of the deuteron nuclear correction is then
$T_i^d[f_d^{(0)},f_p^{(0)}]-T_i^d[f_s^{(0)},f_p^{(0)}]$. However we must also include
the uncertainty in the proton fit, as part of the theoretical (proton)
uncertainty in determining the deuteron PDF from these data: this can be done
by computing the theory covariance matrix 
\begin{equation}
  S_{ij}^p
  =
  \langle\Delta_i^{p,(k)}\Delta_j^{p,(k)}\rangle\,,
  \qquad\qquad
  \Delta_i^{p,(k)}=T_i^d[f_d^{(0)},f_p^{(k)}]-T_i^d[f_d^{(0)},f_p^{(0)}]\,,
  \label{eq:protoncovmatrix}
\end{equation}
where $i,j$ run over the data points in the mixed deuteron datasets only. Note
that this covariance matrix incorporates correlations between the mixed
datasets due to their common dependence on the proton PDF. This theoretical
covariance matrix is added to the experimental covariance matrix of the mixed
deuteron data when performing the deuteron fit, to take account of the
uncertainty of the proton PDF in the determination of the deuteron PDFs.

Note that the theory covariance matrix Eq.~\eqref{eq:protoncovmatrix} itself
depends on the deuteron PDF, which is what we are trying to determine. However 
the dependence of the fitted deuteron PDF on the uncertainty in the proton PDF
is relatively weak, since it only affects the weight of the mixed data in the
fit. Thus to a good approximation we can replace $f_d^{(0)}$ in
Eq.~\eqref{eq:protoncovmatrix} with $f_s^{(0)}$, determined from the pure
proton fit. For a more accurate determination of the deuteron PDFs, we could
then iterate to consistency, performing a second fit to the deuteron data where
$f_d^{(0)}$ in Eq.~\eqref{eq:protoncovmatrix} is determined from the first fit.
It is clear that this iterative process would converge very rapidly. 

However our aim here is not so much to determine the deuteron PDF, but rather
to use it to determine a theoretical covariance matrix that takes into account
the nuclear effects in the deuteron data (both pure and mixed) when using these
data in a global fit of the proton PDF. Since the size of the nuclear correction
is given by the difference between predictions with deuteron and isosinglet
PDFs, this theoretical (deuteron) covariance matrix is
\begin{equation}
  S_{ij}^d
  =\langle\Delta_i^{d,(k)}\Delta_j^{d,(k)}\rangle\,,
  \qquad\qquad
  \Delta_i^{d,(k)}= \left\{\begin{array}{r@{\qquad}l}
    T_i^d[f_d^{(k)}]-T_i^d[f_s^{(0)}]  & i\in {\rm pure} \\[1ex]
    T_i^d[f_d^{(k)},f_p^{(0)}]-T_i^d[f_s^{(0)},f_p^{(0)}]  &  i\in {\rm mixed.}
   \end{array}\right.\,
  \label{eq:deuteroncovmatrix}
\end{equation}
Again this covariance matrix incorporates correlations between all the nuclear
corrections in the various deuteron datasets, due to their common dependence on
the deuteron PDF. To perform a global fit of the proton PDF including
nuclear uncertainties in the deuteron data, we can simply add the theory
covariance matrix to the experimental covariance matrix of the deuteron
datasets and perform the proton fit in the usual way; this yields a set of
replicas of proton PDFs $\{f_p^{(k)}\}$.

Since the global proton PDFs will be more precise than the pure proton PDFs we
started with, it makes sense once again to iterate; we use our global proton
PDFs to determine an improved theoretical proton covariance matrix
Eq.~\eqref{eq:protoncovmatrix}, repeat the deuteron fit, use this to determine
an improved theoretical deuteron covariance matrix
Eq.~\eqref{eq:deuteroncovmatrix}, and then use this to perform a new global
fit of the proton PDF. This is the iterative procedure shown schematically in
Fig.~\ref{fig:procedure}. Note that through this procedure the deuteron PDF is
also iterated concurrently. We expect the iterations to converge very rapidly
to a self consistent set of deuteron and (global) proton PDFs for several
reasons: firstly, a small change in the proton PDF makes a small difference to
the deuteron correction; secondly, we expect the effect of the deuteron
correction on the weight of these data in the global fit to be small; thirdly,
the influence of the deuteron data in the global fit is already relatively
small (just as the influence of the proton PDF on the deuteron fit is small).
Note that the deuteron data are not double counted in this procedure; in the
deuteron fit they are used to determine (empirically) the nuclear uncertainty,
while in the global fit they influence the central value of the proton PDF
directly, but taking into account this nuclear uncertainty. Indeed, the nuclear
uncertainty reduces the weight of the deuteron data in the global fit, so they
actually count less. As a byproduct, we also determine a set of deuteron PDFs.

This realises the first of the two procedures described in~\cite{Ball:2018twp},
whereby the deuteron datasets are deweighted by the nuclear uncertainty but
theoretical predictions are not shifted by a nuclear correction.
As in~\cite{Ball:2018twp}, we also implement the second procedure:
in this case, the theoretical (deuteron) covariance matrix is defined as
\begin{equation}
  S_{ij}^d
  =\langle\Delta_i^{d,(k)}\Delta_j^{d,(k)}\rangle\,,
  \qquad\qquad
  \Delta_i^{d,(k)}= \left\{\begin{array}{r@{\qquad}l}
    T_i^d[f_d^{(k)}]-T_i^d[f_d^{(0)}]  & i\in {\rm pure} \\[1ex]
    T_i^d[f_d^{(k)},f_p^{(0)}]-T_i^d[f_d^{(0)},f_p^{(0)}]  &  i\in {\rm mixed,}
   \end{array}\right.
  \label{eq:deuteroncovmatrixnew}
\end{equation}
while the corrections applied to the theoretical predictions $T_i^d[f_s^{(0)}]$
and $T_i^d[f_s^{(0)},f_p^{(0)}]$ for the deuteron datasets are
\begin{equation}
  \delta T_i^d=\left\{\begin{array}{r@{\qquad}l}
    T_i^d[f_d^{(0)}]-T_i^d[f_s^{(0)}]  & i\in {\rm pure} \\[1ex]
    T_i^d[f_d^{(0)},f_p^{(0)}]-T_i^d[f_s^{(0)},f_p^{(0)}]  &  i\in {\rm mixed.}
   \end{array}\right. 
  \label{eq:deuteroncovmatrixshift}
\end{equation}
This procedure is implemented in the same way as the first, with the
theoretical predictions for the deuteron data corrected before performing each
iteration of the global proton fit. Since, unlike in~\cite{Ball:2018twp},
the corrections are determined empirically and self consistently, we expect the
second method to be more precise than the first method; the central
values of the theoretical predictions should be a little more accurate, and the
uncertainty due to nuclear effects in the deuteron correspondingly a little
smaller.

\begin{table}[!t]
  \rowcolors{2}{gray!25}{white}
  \centering
  \scriptsize
  \renewcommand{\arraystretch}{1.13}
  \begin{tabularx}{\textwidth}{lllX}
    \toprule
    \rowcolor{gray!40}
    {\bf Iteration }  & {\bf Dataset }            & {\bf Fit ID }        & { \bf Description }\\
    \midrule
    Baseline    & Proton and Deuteron & global-base    & Same as {\tt base} fit in~\cite{Faura:2020oom}.\\
    \midrule
    Iteration 0 & Proton              & proton-ite0    & Same as baseline, but restricted to the proton dataset\\
    \midrule
    Iteration 1 & Deuteron            & deuteron-ite1  & Same as baseline, but restricted to the deuteron dataset
                                                         and supplemented with a proton covariance matrix determined
                                                         from the proton-ite0 fit according to
                                                         Eq.~\eqref{eq:protoncovmatrix}.\\
                & Proton and Deuteron & global-ite1-dw & Same as baseline, but supplemented with a deuteron covariance
                                                         matrix determined from the deuteron-ite1 fit according to
                                                         Eq.~\eqref{eq:deuteroncovmatrix}.\\
    \midrule
    Iteration 2 & Deuteron            & deuteron-ite2  & Same as deuteron-ite1, but with a proton covariance matrix
                                                         determined from the global-ite1-dw fit.\\
                & Proton and Deuteron & global-ite2-dw & Same as global-ite1-dw, but with a deuteron covariance matrix
                                                         determined from the deuteron-ite2 fit.\\
                & Proton and Deuteron & global-ite2-sh & Same as global-ite2-sh, but with a deuteron covariance matrix
                                                         and shifts determined according to
                                                         Eqs.~\eqref{eq:deuteroncovmatrixnew}-\eqref{eq:deuteroncovmatrixshift}.\\
    \bottomrule
  \end{tabularx}
  \caption{A summary of the fits performed in this study, see text for details.}
  \label{tab:fits}
\end{table}


The set of fits which we performed, all using NNPDF methodology, are summarised
in Table~\ref{tab:fits}. They are all accurate to next-to-next-to-leading order
(NNLO) in perturbative QCD, heavy quarks are treated in the FONLL scheme and
the charm PDF is parametrised in the same way as the lighter quark
PDFs. All PDF sets are made of $N_{\rm rep}=100$ Monte Carlo replicas.
The baseline fit, `global-base', is a fit equivalent to the {\tt base} fit
performed in~\cite{Faura:2020oom}.
It is a minor variant of the fit presented in~\cite{Ball:2018iqk}: a bug
affecting the computation of theoretical predictions for charged-current DIS
cross sections has been corrected; positivity of the $F_2^c$ structure
function has been enforced; and NNLO massive
corrections~\cite{Berger:2016inr,Gao:2017kkx}
have been included in the computation of neutrino-DIS structure functions. 
`proton-ite0' is the corresponding fit based on the proton data alone.
We then perform two iterations of the procedure
described above (denoted as iteration 1 and 2), after which we determine a
fit of deuteron PDFs (based only on the deuteron data, and
supplemented with a proton covariance matrix), and a global fit of proton PDFs
(based on the proton and deuteron data, and supplemented with a
deuteron covariance matrix). The deuteron fits 
`deuteron-ite1' and `deuteron-ite2', are performed using exactly the same
theoretical and methodological settings as the proton fits, except that the
isotriplet PDFs are set to zero since the deuteron is isoscalar. After the first
iteration we produce a single global fit of proton PDFs, `global-ite1-dw',
in which the deuteron covariance matrix is evaluated according to
Eq.~\eqref{eq:deuteroncovmatrix}. After the second iteration we produce instead
two global fits of proton PDFs: `global-ite2-dw' in which the deuteron
covariance matrix
is evaluated with Eq.~\eqref{eq:deuteroncovmatrix}, and `global-ite2-sh',
in which the deuteron covariance matrix is evaluated with
Eq.~\eqref{eq:deuteroncovmatrixnew}, and the theoretical predictions are 
first corrected according to Eq.~\eqref{eq:deuteroncovmatrixshift}.

Before discussing the results of our fits, we look more closely at the pattern
of deuteron corrections, and at the deuteron covariance matrix defined in
Eq.~\eqref{eq:deuteroncovmatrix}. As representative examples, results are
obtained from proton and deuteron PDFs determined after the first iteration.
We explicitly checked that they remain stable after an additional iteration.

\begin{figure}[!t]
  \centering
  \includegraphics[width=\linewidth,clip=true,trim=4cm 0 4cm 0]{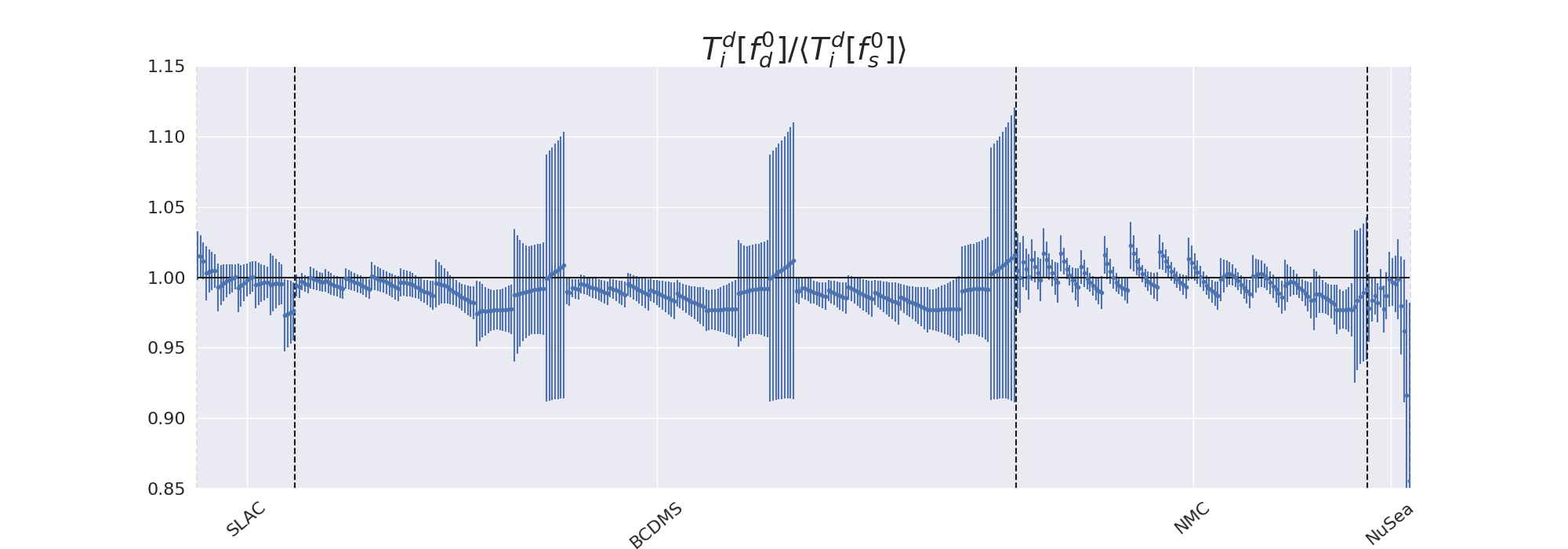}\\
  \caption{The ratio between the SLAC, BCDMS, NMC and DYE866/NuSea deuteron
    observables computed either with the central prediction with deuteron PDFs,
    $T_i^d[f^0_d]$, or the central prediction with proton PDFs, $T_i^d[f^0_s]$.
    Data points are ordered in bins of increasing values of momentum fraction
    $x$ and energy $Q$.}
  \label{fig:ratio}
\end{figure}

In Fig.~\ref{fig:ratio} we display the nuclear correction for the deuteron data
obtained from our procedure. Specifically, for each data point $i$ (after
kinematic cuts), we show the observables computed with the central deuteron PDF,
normalised to the expectation value computed with the central proton PDF,
$T_i^d[f^0_d]/ T_i^d[f^0_s]$. Data points are ordered in bins of
increasing values of momentum fraction $x$ and energy $Q$. The deuteron
correction generally amounts to a few percent for all of the
experiments considered. Uncertainties are rather large and the
ratio is mostly compatible with one, except for data points at higher values
of momentum fraction $x$ and energy $Q$, where the correction is negative, as 
expected from models of nuclear shadowing.

\begin{figure}[!t]
  \centering
   \includegraphics[width=\linewidth,clip=true,trim=4cm 0 4cm 0]{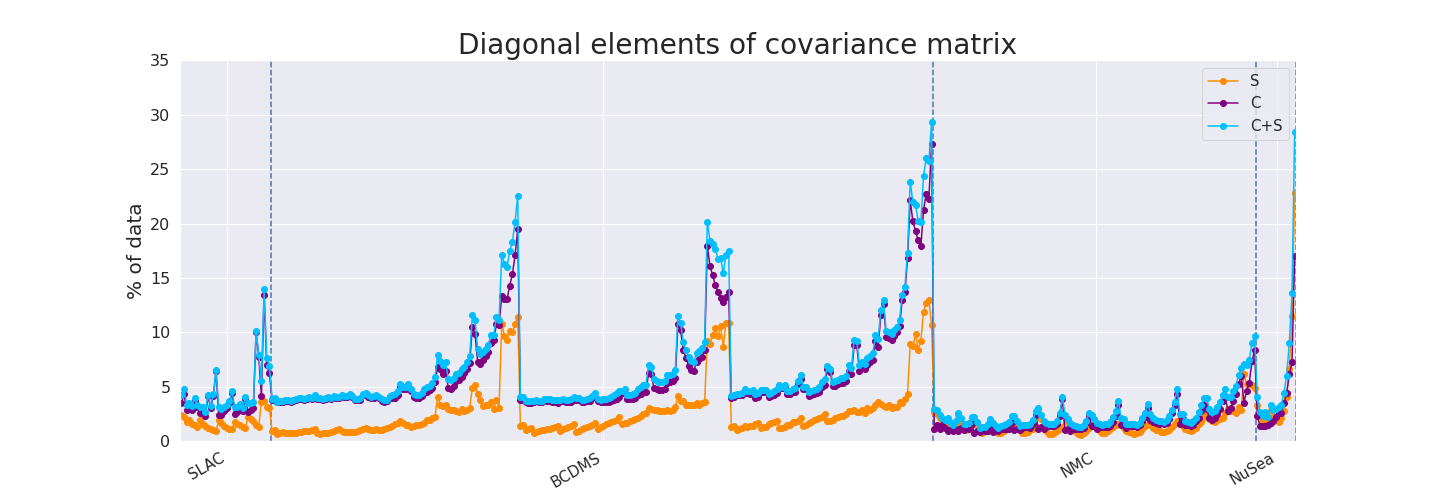}\\
   \caption{The square root of the diagonal elements of the covariance matrices
     normalised to the experimental data, $\sqrt{{\rm cov}_{ii}}/D_i$, for the
     deuteron measurements from SLAC, BCDMS, NMC and DYE866/NuSea. We show
     results for the experimental covaraince matrix $(C)$, for the deuteron
     covariance matrix $(S)$, computed from Eq.~\eqref{eq:deuteroncovmatrix},
     and for their sum $(C+S)$.}
  \label{fig:diagonal}
\end{figure}

To gain a further idea of the effects to be expected from nuclear corrections
in deuteron, in Fig.~\ref{fig:diagonal} we show the square root of the diagonal
elements of the experimental $(C)$ and theoretical $(S)$ covariance matrices,
and their sum $(C+S)$, each normalised to the central value of the experimental
data: $\sqrt{{\rm cov}_{ii}}/D_i$. The theoretical covariance matrix
accounts for the nuclear uncertainties, and is computed with
Eq.~\eqref{eq:deuteroncovmatrix}. The general pattern of the results does not
change qualitatively if
Eqs.~\eqref{eq:deuteroncovmatrixnew}-\eqref{eq:deuteroncovmatrixshift}
are used instead.
The pattern observed in Fig.~\ref{fig:ratio} is paralleled in
Fig.~\ref{fig:diagonal}, in particular concerning the dependence of the size of
the nuclear uncertainties on the bin kinematics for each experiment. Moreover
we can now see that the deuteron uncertainties are smaller than the data
uncertainties for SLAC and BCDMS, while they are comparable for NMC and
DYE866/NuSea. This is largely because the pure deuteron measurements from
SLAC and BCDMS are of cross-sections, whereas the more precise mixed
measurements from NMC and DYE866/NuSea 
are of cross-section ratios, for which systematic uncertainties largely cancel. 

\begin{figure}[!t]
  \centering
  \includegraphics[width=0.47\textwidth]{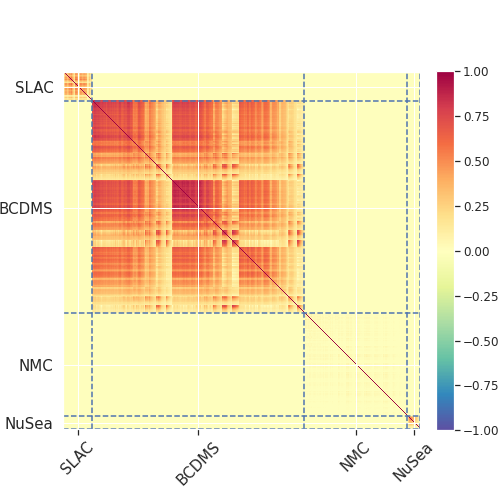} \ \ \ \ \
  \includegraphics[width=0.47\textwidth]{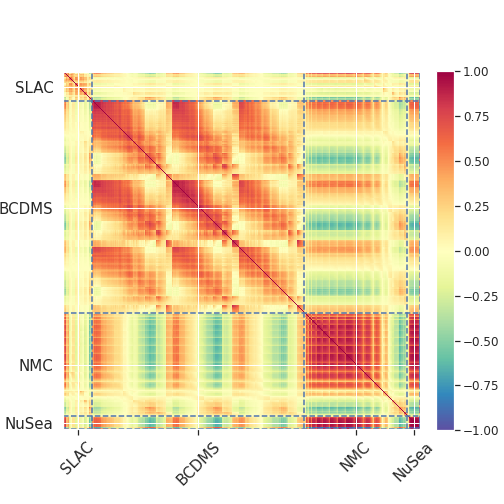}
  \caption{The experimental (left) and total (right) correlation matrices for
    the SLAC, BCDMS, NMC and DYE866/NuSea deuteron experiments. The deuteron
    covariance matrix, added to the experimental covariance matrix to obtain
    the total covariance matrix, is computed according to
    Eq.~\eqref{eq:deuteroncovmatrix}.}
  \label{fig:correlation}
\end{figure}

Finally, we show the experimental correlation
matrix, $\rho_{ij}^C=C_{ij}/\sqrt{C_{ii}C_{jj}}$, and the sum of the experimental
and deuteron correlation matrices,
$\rho_{ij}^{C+S}=(C_{ij}+S_{ij})/\sqrt{(C_{ii}+S_{ii})(C_{jj}+S_{jj})}$, as heat
plots in Fig.~\ref{fig:correlation}.
Note that, whenever the proton and deuteron data are both included in the
global proton fit, there are small normalisation uncertainties correlated
between $F_2^p$ and $F_2^d$ measurements within the SLAC and BCDMS experiments.
These correlations, not shown in Fig.~\ref{fig:correlation}, are taken into
account by default in all NNPDF analyses, including this one (for details,
see Sect.~2.1 in~\cite{Forte:2002fg}).
The theoretical covariance matrix is computed according to
Eq.~\eqref{eq:deuteroncovmatrix}, though the
qualitative behaviour of the total correlation matrix is unaltered if
Eqs.~\eqref{eq:deuteroncovmatrixnew}-\eqref{eq:deuteroncovmatrixshift} are
used instead. Our procedure captures
the sizeable correlations of the deuteron corrections between different
bins of momentum and energy, systematically enhancing bin-by-bin (positive and
negative) correlations in the data. As we might expect, nuclear
uncertainties are also strongly correlated across the different experiments.

We now turn to discuss the results of the fits collected in
Table~\ref{tab:fits}. In Tables~\ref{tab:chi2-deut}-\ref{tab:chi2-global}
we display the values of the experimental $\chi^2$ per data point (as defined
in Eq.~(4) of~\cite{Ball:2018twp}) for the fits of the deuteron PDFs (based on
the deuteron data) and of the proton PDFs (based on the global dataset
including both deuteron and proton data), respectively.
In Table~\ref{tab:chi2-global} values are displayed both for separate
datasets, and for groups of datasets corresponding to measurements of similar
observables in the same experiment. Indented datasets are subsets of the
preceding non-indented dataset.

\begin{table}[!t]
  \rowcolors{2}{gray!25}{white}
  \centering
  \scriptsize
  \renewcommand{\arraystretch}{1.13}
  \begin{tabularx}{\textwidth}{lCCC}
    \toprule
    \rowcolor{gray!40}
    \textbf{Experiment}
    & \textbf{$N_{\rm dat}$}
    &\textbf{ deuteron-ite1}
    & \textbf{deuteron-ite2}\\
    \midrule
    SLAC  ($F_2^d$)
    &      34   &      0.81   &      0.84   \\
    BCDMS ($F_2^d$)
    &     248   &      1.02   &      1.04   \\
    NMC   ($F_2^d/F_2^p$)
    &     121   &      1.02   &      0.99   \\
    DYE866/NuSea ($\sigma_{pd}^{\rm DY}/\sigma_{pp}^{\rm DY}$)
    &      15   &      0.14   &      0.14   \\
    \midrule
    \rowcolor{gray!40}
    {\bf Total } & {\bf 418} & {\bf 0.97} & {\bf 0.98} \\
        \rowcolor{white}
    \bottomrule\\
  \end{tabularx}
  \caption{The values of the $\chi^2$ per data point for each dataset included
    in the fits of deuteron PDFs.}
  \label{tab:chi2-deut}
\end{table}

\begin{table}[!t]
  \rowcolors{2}{gray!25}{white}
  \centering
  \scriptsize
  \renewcommand{\arraystretch}{1.13}
  \begin{tabularx}{\textwidth}{lCCCCC}
    \toprule 
    \rowcolor{gray!40}
    \textbf{Experiment}
    &  \textbf{$N_{\rm dat}$} &  \textbf{global-base} &  \textbf{global-ite1-dw} &  
    \textbf{global-ite2-dw} &  \textbf{global-ite2-sh} \\
    \midrule
    SLAC  ($F_2^d$)
    &         34 & 0.72 & 0.50 & 0.50 & 0.49 \\
    BCDMS ($F_2^d$)
    &        248 & 1.10 & 0.98 & 0.91 & 0.96 \\
    NMC   ($F_2^d/F_2^p$)
    &        121 & 1.00 & 0.79 & 0.78 & 0.82 \\
    DYE886/NuSea ($\sigma_{pd}^{\rm DY}/\sigma_{pp}^{\rm DY}$)
    &         15 & 0.47 & 0.53 & 0.71 & 1.06 \\
    \midrule
    SLAC  ($F_2^p$)
    &         33 & 0.88 & 0.93 & 0.91 & 0.91 \\
    BCDMS ($F_2^p$)
    &        333 & 1.30 & 1.30 & 1.29 & 1.30 \\
    NMC   ($F_2^p$)
    &        204 & 1.54 & 1.55 & 1.54 & 1.55 \\
    CHORUS
    &        832 & 1.15 & 1.14 & 1.14 & 1.14 \\
    NuTeV
    &         76 & 0.76 & 0.81 & 0.77 & 0.89 \\
    HERA I+II (incl.)
    &       1145 & 1.16 & 1.16 & 1.16 & 1.16 \\
    HERA ($\sigma_c^{\rm NC}$)
    &         37 & 1.42 & 1.43 & 1.43 & 1.44 \\
    HERA ($F_2^b$)
    &         29 & 1.11 & 1.11 & 1.11 & 1.11 \\
    DYE866/NuSea ($\sigma_{p}^{\rm DY}$)
    &         89 & 1.29 & 1.24 & 1.21 & 1.17 \\
    DYE605 ($\sigma_{p}^{\rm DY}$)
    &         85 & 1.11 & 1.10 & 1.09 & 1.12 \\
    CDF ($Z$ rap.)
    &         29 & 1.36 & 1.34 & 1.35 & 1.33 \\
    D0 ($Z$ rap.; $W$ asy.)
    &         45 & 1.18 & 1.15 & 1.16 & 1.17 \\
    ATLAS 
    &        211 & 1.13 & 1.12 & 1.12 & 1.10 \\
    \ \ \ ATLAS (Drell--Yan)
    &         75 & 1.44 & 1.40 & 1.41 & 1.35 \\
    \ \ \ ATLAS (jets)
    &         31 & 1.09 & 1.11 & 1.10 & 1.08 \\
    \ \ \ ATLAS ($Z p_T$)
    &         92 & 0.88 & 0.88 & 0.89 & 0.89 \\
    \ \ \ ATLAS (top)
    &         13 & 1.31 & 1.25 & 1.26 & 1.24 \\
    CMS 
    &        327 & 1.15 & 1.13 & 1.14 & 1.15 \\
    \ \ \ CMS (Drell--Yan)
    &        154 & 1.27 & 1.28 & 1.28 & 1.28 \\
    \ \ \ CMS (jets)
    &        133 & 0.99 & 0.93 & 0.97 & 1.05 \\
    \ \ \ CMS ($Z p_T$)
    &         28 & 1.32 & 1.30 & 1.31 & 1.31 \\
    \ \ \ CMS (top)
    &         12 & 0.88 & 0.87 & 0.89 & 0.87 \\
    LHCb
    &         85 & 1.62 & 1.59 & 1.67 & 1.66 \\
    \midrule
    \rowcolor{gray!40}
    {\bf Total } & {\bf 3978} & {\bf 1.18} & {\bf 1.16} & {\bf 1.16} & {\bf 1.16} \\
    \bottomrule\\
  \end{tabularx}
  \caption{The values of the $\chi^2$ per data point for each dataset included
    in the global fits of proton PDFs. The deuteron data are at the top of the table.}
  \label{tab:chi2-global}
\end{table}

\begin{figure}[!t]
\centering
\includegraphics[width=0.49\linewidth]{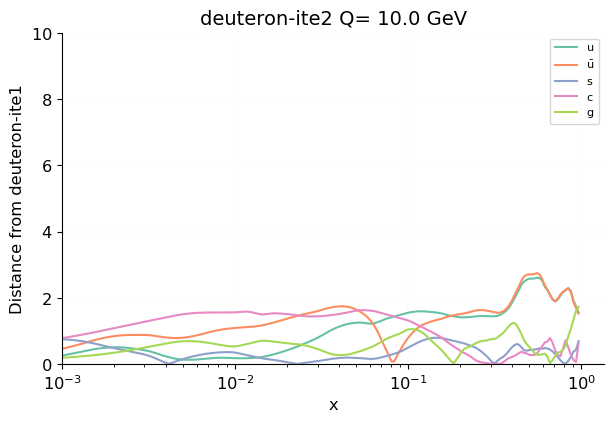}
\includegraphics[width=0.49\linewidth]{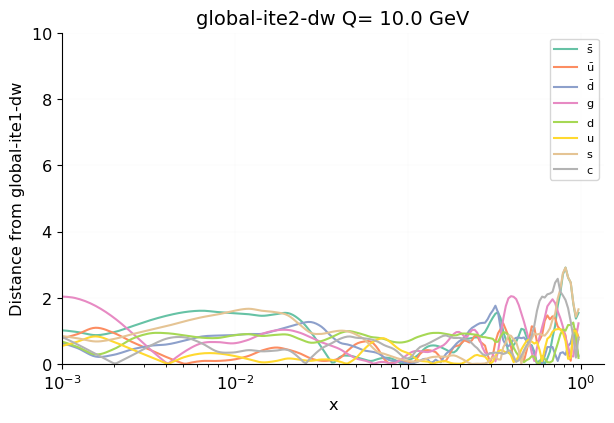}
\caption{Distances between the central values of the deuteron-ite1 and
  deuteron-ite2 fits (left) and of the global-ite1-dw and global-ite2-dw fits
  (right), see Table~\ref{tab:fits} for details. For the deuteron fits, $u$
  and $\bar u$ actually denote the combinations $(u+d)/2$ and
  $(\bar u+\bar d)/2$, where $u=d$ and $\bar u=\bar d$ by definition.
  Results are displayed as a
  function of $x$ at a representative scale of the deuteron dataset,
  $Q=10$~GeV. The {\sc ReportEngine}
  software~\cite{zahari_kassabov_2019_2571601} was used to generate this
  figure.}
\label{fig:distances_convergence}
\end{figure}

In order to examine the convergence of our procedure, we must quantifiy the
statistical equivalence between pairs of PDFs obtained from the various fits.
To this purpose we display in Fig.~\ref{fig:distances_convergence} the distance 
(as defined in Eq.~(63) of~\cite{Ball:2010de}) between the central values of
the two iterations of fits based on the deuteron data (deuteron-ite1 and
deuteron-ite2), and the corresponding two iterations on the global data
(global-ite1-dw and global-ite2-dw). For two PDF sets made of $N_{\rm rep}=100$
replicas, a distance of $d\simeq 1$ corresponds to statistically equivalent
sets, while a distance of $d\simeq 10$ corresponds to sets that differ by one
sigma in units of the corresponding standard deviation. Note that in the left
panel of Fig.~\ref{fig:distances_convergence} $u$ and $\bar u$ actually denote
the combinations $(u+d)/2$ and $(\bar u+\bar d)/2$, where $u=d$ and
$\bar u=\bar d$ by definition. These are the isosinglet combinations determined
in the fits to deuteron data.

From the results displayed in Tables~\ref{tab:chi2-deut}-\ref{tab:chi2-global}
and in Fig.~\ref{fig:distances_convergence}, we can conclude that one iteration
is sufficient to achieve stability. The variation of the global $\chi^2$ per
data point for fits obtained in subsequent iterations is smaller than
statistical fluctuations. This is true both in the case of fits of deuteron PDFs
(the global $\chi^2$ per data point is 0.97 and 0.98 for the deuteron-ite1 and
deuteron-ite2 fits, respectively), and in the case of global fits of proton
PDFs (the global $\chi^2$ per data point is 1.16 in both the global-ite1-dw and
global-ite2-dw fits), see Tables~\ref{tab:chi2-deut}-\ref{tab:chi2-global}.
Variations of the $\chi^2$ per data point for single experiments are likewise
smaller than statistical fluctuations. Furthermore, distances between the
central values of the corresponding PDFs are at most of the order of two or
three, for both deuteron and proton fits. The PDF flavours that change the most
upon iteration are $u$, $\bar{u}$, $d$, $\bar{d}$ in the valence region of the
deuteron fit, as might be expected. 

\begin{figure}[!t]
\centering
\includegraphics[width=0.49\linewidth]{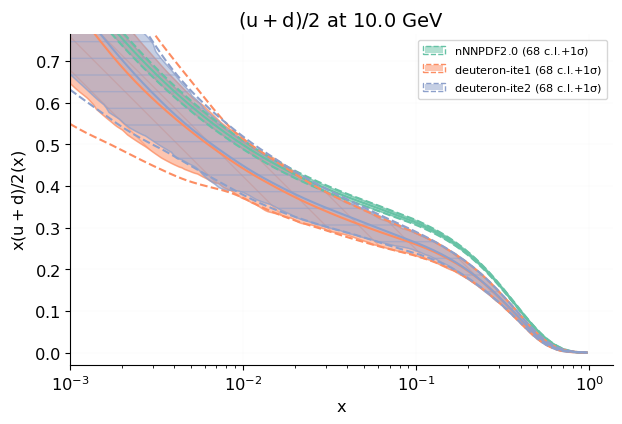}
\includegraphics[width=0.49\linewidth]{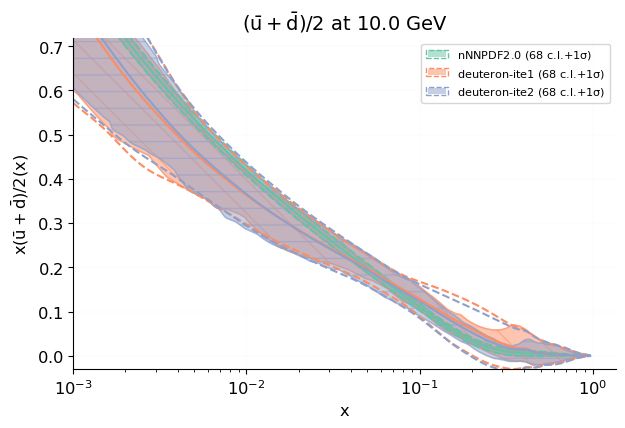}\\
\includegraphics[width=0.49\linewidth]{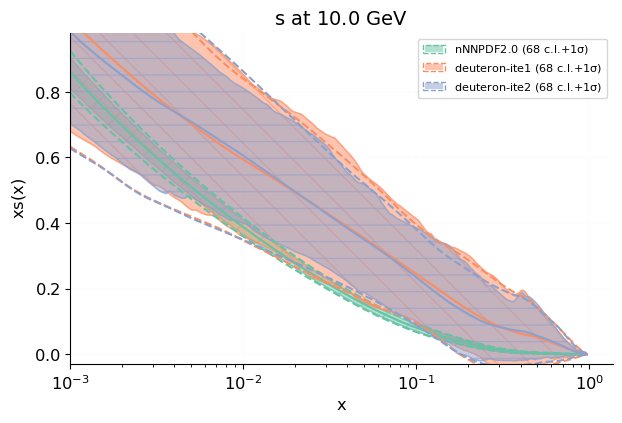}
\includegraphics[width=0.49\linewidth]{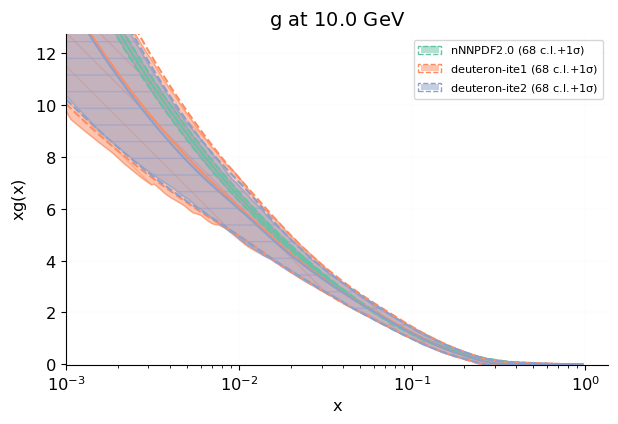}\\
\caption{Comparison between the deuteron-ite1, deuteron-ite2 and nNNPDF2.0~\cite{AbdulKhalek:2020yuc}
  deuteron PDFs. The average of up and down, the average of antiup and antidown,
  strange and gluon PDFs are shown at $Q=10$~GeV.  Dashed lines denote one
  sigma uncertainties, while plain bands 68\% confidence level intervals.
  The {\sc ReportEngine} software~\cite{zahari_kassabov_2019_2571601}
  was used to generate this figure.}
\label{fig:deuteron_pdfs}
\end{figure}

Next, in Fig.~\ref{fig:deuteron_pdfs} we compare the
deuteron PDFs obtained from each iteration (deuteron-ite1 and deuteron-ite2).
Specifically we show the average of up and down quark, the average of up
and down antiquark, the strange quark, and the gluon
distributions: because the deuteron is isoscalar, $d=u$ and $\bar{d}=\bar{u}$
by construction. Again we see that the PDFs hardly
change from one iteration to the next, so the procedure has converged.
In addition in this plot we compare our NNLO deuteron PDFs with a recent NLO
determination of nuclear PDFs based on the NNPDF methodology,
nNNPDF2.0~\cite{AbdulKhalek:2020yuc}. These were obtained by 
fitting a range of nuclear and proton data, and assuming a smooth dependence
on the mass and atomic numbers $A$ and $Z$. Due to this assumption, 
which in effect constrains the deuteron as an interpolation between proton and
heavy nuclei, their uncertainties are smaller than our own.
Our determination of the deuteron PDFs is thus very conservative.

Note that the central values of the deuteron PDFs in nNNPDF2.0 are mostly
consistent with ours within 
uncertainties. A discrepancy of about one sigma, in units of the uncertainty of
the deuteron-ite2 fit, is observed for the average of up and down quarks
around $x\sim 0.1$. Whether this discrepancy might be explained in light of the
fact that the two determinations are at different orders in perturbation
theory remains unclear. Available nuclear PDF sets accurate to
NNLO~\cite{AbdulKhalek:2019mzd,Walt:2019slu} currently include only inclusive
DIS measurements, and thus have larger PDF uncertainties than nNNPDF2.0.
This obscures the phenomenological impact of higher order corrections.

\begin{figure}[!t]
\centering
\includegraphics[width=0.5\linewidth]{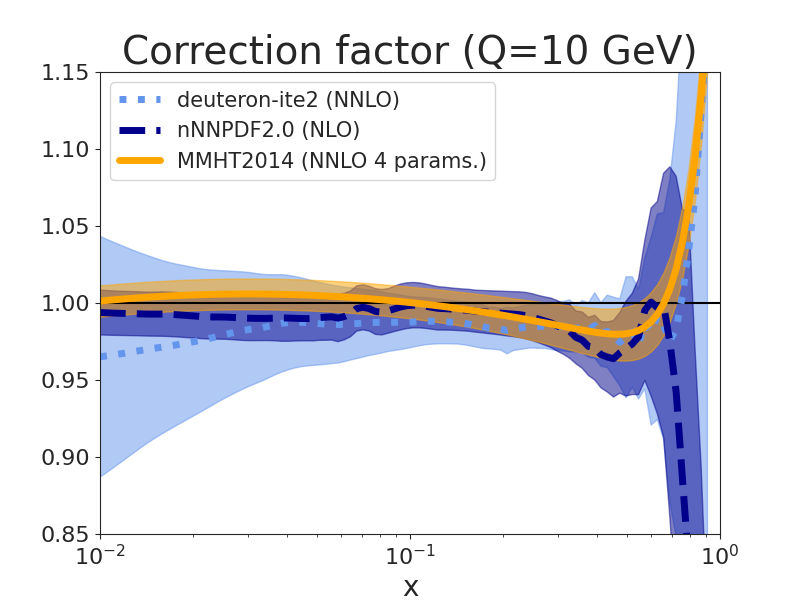}
\caption{The nuclear correction factor $F_2^d/F_2^p$, calculated using our final
  deuteron fit deuteron-ite2, the deuteron PDF from nNNPDF2.0, and the model
  fit used for deuteron corrections in MMHT2014. Results are displayed as a
  function of $x$ at the representative scale $Q=10$~GeV.}
\label{fig:nuclear_corrn}
\end{figure}

To explore the nuclear corrections further, in Fig.~\ref{fig:nuclear_corrn}
we show the ratio $F_2^d/F_2^p$ computed with our deuteron  PDFs, at $Q=10$~GeV.
We see that the correction for nuclear effects in deuteron is only a few per
cent over the full range of $x$, and is negative in the valence region, as 
expected from nuclear shadowing. However the uncertainty in our determination
is as large as the correction. For comparison, we also show the same 
quantity computed using the nNNPDF2.0 deuteron PDFs~\cite{AbdulKhalek:2020yuc}:
these are NLO, but have a smaller uncertainty since, as explained above, in
these fits continuity in $A/Z$ is implicitly assumed, which adds a significant
constraint. However the reduction in uncertainty due to this constraint is
considerably less in the structure function ratio than it was in the PDFs.
We also show the parametric correction used in the MMHT14
fits~\cite{Harland-Lang:2014zoa}, which has four fitted parameters.
Again this has a yet smaller uncertainty, particularly at large $x$, due to the
assumed theoretical constraints of the model. However we note that all three of
these estimates are mutually consistent, within uncertainties. The estimate
obtained here is clearly the most conservative, particularly outside the
valence region, as expected since it is free from any model dependence. 

\begin{figure}[!t]
\centering
\includegraphics[width=0.49\linewidth]{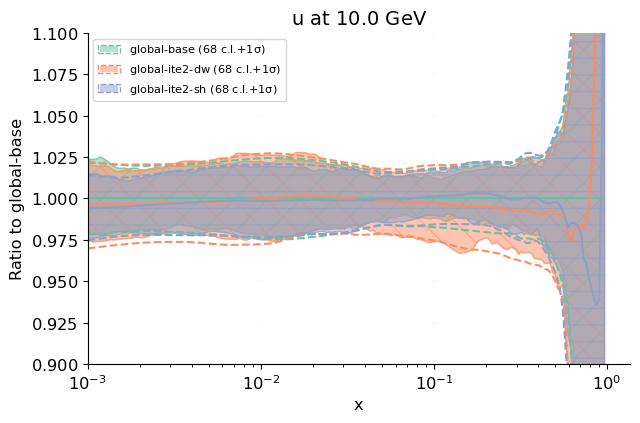}
\includegraphics[width=0.49\linewidth]{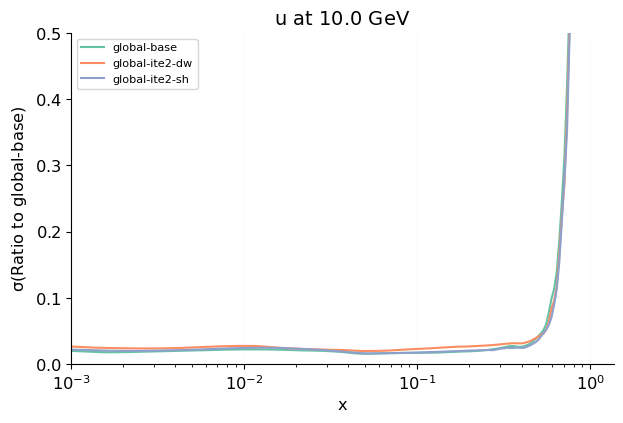}\\
\includegraphics[width=0.49\linewidth]{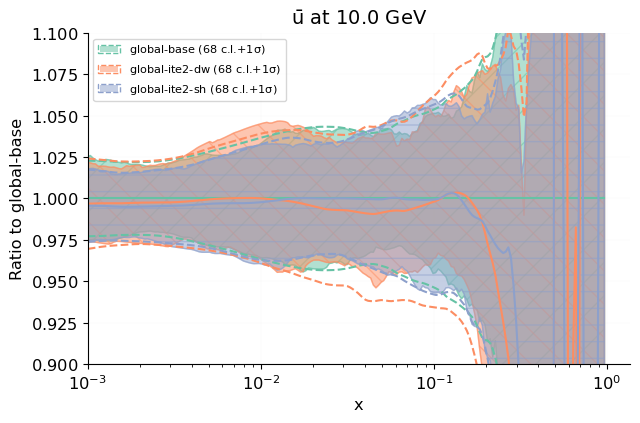}
\includegraphics[width=0.49\linewidth]{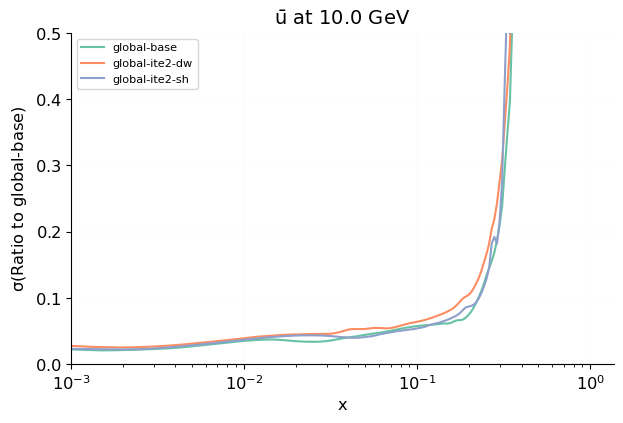}\\
\includegraphics[width=0.49\linewidth]{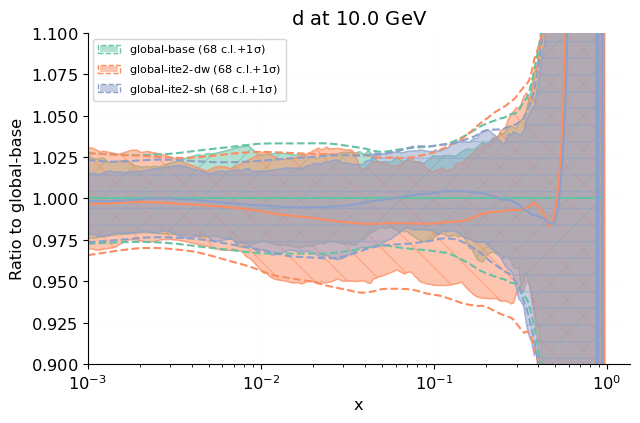}
\includegraphics[width=0.49\linewidth]{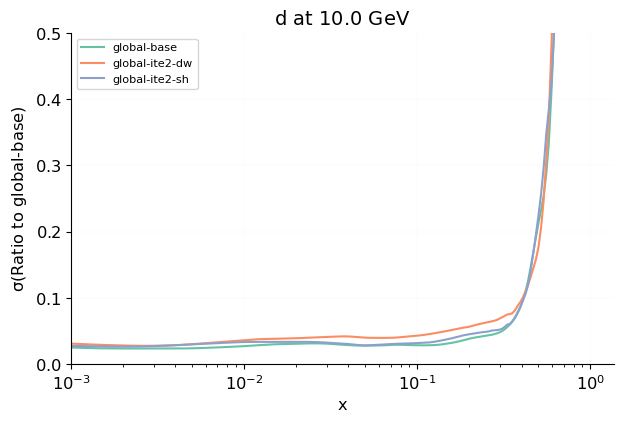}\\
\includegraphics[width=0.49\linewidth]{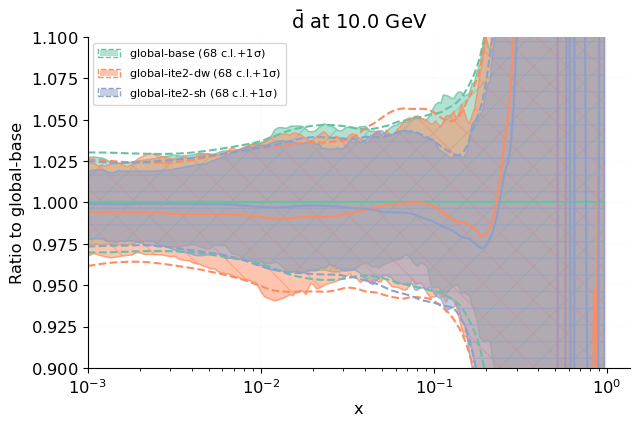}
\includegraphics[width=0.49\linewidth]{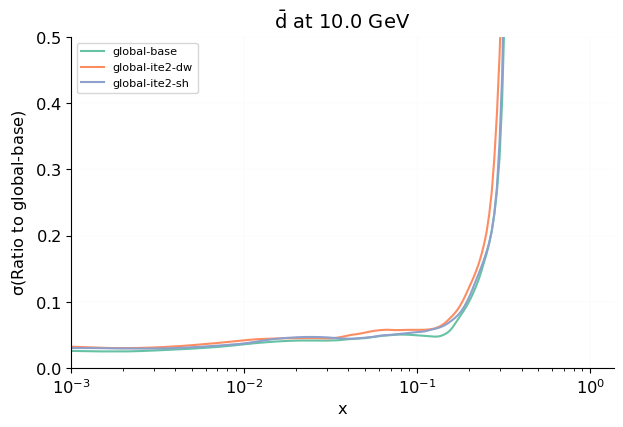}\\
\caption{Comparison between the global-base, global-ite2-dw and global-ite2-sh
  global fits of proton PDFs. The up, antiup, down and antidown PDFs,
  normalised to the global-base fit (left) and the corresponding relative
  uncertainties (right) are shown at $Q=10$~GeV.  Dashed lines denote
one sigma uncertainties, while plain bands 68\% confidence level intervals. 
The {\sc ReportEngine} software~\cite{zahari_kassabov_2019_2571601} was used to
generate this figure.}
\label{fig:proton_pdfs}
\end{figure}

\begin{figure}[!t]
\centering
\includegraphics[width=0.49\linewidth]{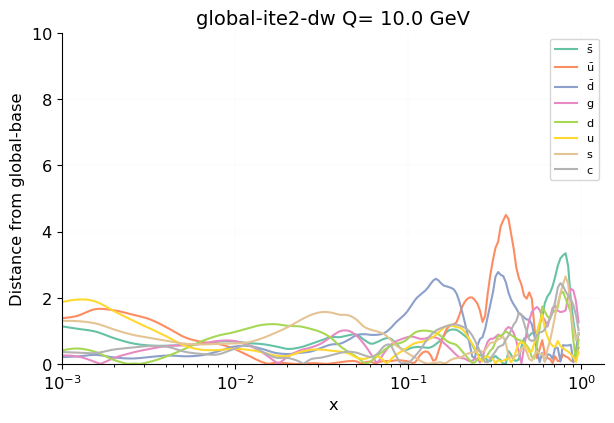}
\includegraphics[width=0.49\linewidth]{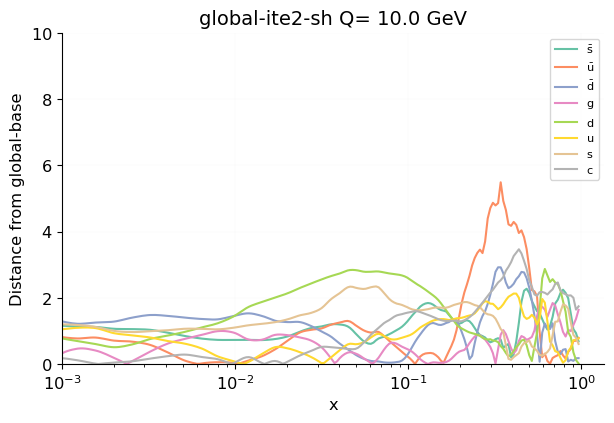}\\
\caption{Distances between the central values of the global-base and
  global-ite2-dw fits (left) and of the global-base and global-ite2-sh fits
  (right), see Table~\ref{tab:fits} for details. Results are displayed as a
  function of $x$ at a representative scale for the deuteron dataset,
  $Q=10$~GeV. The {\sc ReportEngine} software~\cite{zahari_kassabov_2019_2571601} was used to
generate this figure.}
\label{fig:distances_impact}
\end{figure}

Finally we consider the impact of the deuteron uncertainties in the global fit
of proton PDFs. We compare the global fits (to the deuteron and proton data),
made without the inclusion of the theoretical covariance matrix (global-base),
and then with the inclusion of the theoretical covariance matrix after the
second iteration, either with Eq.~\eqref{eq:deuteroncovmatrix} (global-ite2-dw)
or with Eq.~\eqref{eq:deuteroncovmatrixnew} and the associated shifts,
\eqref{eq:deuteroncovmatrixshift} (global-ite2-sh).
From Table~\ref{tab:chi2-global} we conclude that the nuclear corrections give a
small improvement in the overall fit quality (the global $\chi^2$ per data
point is reduced from 1.18 in the global-base fit to 1.16 in the
global-ite2-dw and global-ite2-sh fits, which corresponds to one standard
deviation of the $\chi^2$ distribution), and a significant improvement in the
fit quality of the deuteron datasets. 

Turning to the PDFs themselves, in Fig.~\ref{fig:proton_pdfs} we compare the
proton PDFs obtained from these three fits. Here we show only the up and down
quark and antiquark PDFs, normalised to the global-base fit, and the
corresponding relative uncertainties, since the other quark flavours and the
gluon PDFs are only scarcely affected by the nuclear corrections in deuteron.
We also show, in Fig.~\ref{fig:distances_impact} the distances (defined as in
Fig.~\ref{fig:distances_convergence}) between the baseline fit (global-base)
and each of the two global fits with deuteron uncertainties included after the
second iteration (global-ite2-dw and global-ite2-sh). Again, a distance of
$d\simeq 10$ corresponds to sets that differ by one sigma in units of the
corresponding standard deviation. 

The effect of the nuclear corrections on the PDF central values is largest in
the up antiquark in the valence region: it differs by about half a sigma
($d\sim 5$ in Fig.~\ref{fig:distances_impact}) in both the global-ite2-dw and
global-ite2-sh fits with respect to the global-base fit. As apparent from
Fig.~\ref{fig:proton_pdfs}, the central value of the up antiquark PDFs is
suppressed in the valence region, while that of the down antiquark is
enhanced. The effect is seen irrespective of whether the theoretical
predictions are shifted. The inclusion of the nuclear uncertainty in the global
fits of proton PDFs results in a slight increase in the uncertainties in
comparison to the global-base fit, but this increase is rather larger in the
global-ite2-dw fit than in the global-ite2-sh fit, where it is scarcely visible.
This result, combined with the fact that both these fits have comparable
quality (see Table~\ref{tab:chi2-global}), leads us to conclude that the
shifted fit is to be preferred. This is as expected, given that the uncertainty
due to nuclear corrections has been determined self-consistently, and
turns out to be a little smaller in the valence region than the nuclear
correction itself (see Fig.~\ref{fig:ratio}). This in contrast to the result we
found in the case of heavy nuclei~\cite{Ball:2018twp}, for which nuclear
uncertainties were instead estimated from independent global determinations of
nuclear PDFs. Clearly the self-consistency of our procedure is advantageous,
and should therefore be preferred in the case of deuteron data (and for heavy
nuclei whenever it is possible to perform a consistently reliable
determination of the nuclear PDFs and their uncertainties).

In summary, we have developed an iterative procedure to incorporate theoretical
uncertainties due to nuclear effects self-consistently into global fits of
proton PDFs that include DIS and DY data on deuterium targets, without
any model dependent assumptions regarding the physics of the nuclear
corrections. In the framework of the NNPDF3.1 global analysis we have shown
that the effect of the additional uncertainty in the global determination of
the proton PDFs is small, and can be reduced further by applying an empirical
correction to the theoretical predictions of the deuteron data. 
Such a fit thus leads to slightly more precise PDFs. We therefore
conclude that, in a fit of proton PDFs including deuteron data,
the approach in which nuclear effects give a correction plus uncertainty
is preferred to the more conservative one in which they give a (larger)
uncertainty only. A similar procedure might be used to improve
the determination of kaon fragmentation functions, and thus the strange and
anti-strange proton PDFs, by means of semi-inclusive DIS measurements that are
sensitive to both. The PDF sets discussed in this work are available in the
{\sc LHAPDF} format~\cite{Buckley:2014ana} from the authors upon request.

\vspace{0.2cm}

\noindent{\bf Acknowledgements.} We thank our colleagues in the NNPDF
collaboration for comments on the manuscript, in particular Rabah Abdul Khalek,
Stefano Forte, Zahari Kassabov and Juan Rojo.
R.D.B. and E.R.N. are supported by the UK STFC
grants ST/P000630/1 and ST/T000600/1. E.R.N. was also supported by the European
Commission through the Marie Sk\l odowska-Curie Action ParDHonSFFs.TMDs
(grant number 752748). R.L.P. is supported by the UK STFC grant ST/R504737/1.

\bibliography{deut-corrs}
\end{document}